# Phase Locking Value revisited: teaching new tricks to an old dog

Ricardo Bruña[1,2], Fernando Maestú[1,2], and Ernesto Pereda[1,3]

[1] Laboratory for Cognitive and Computational Neuroscience, Canter for Biomedical Technology, Technical University of Madrid, Pozuelo de Alarcón, Madrid, Spain

[2] Departamento de Psicología Experimental, Procesos Psicológicos y Logopedia, Faculty of Psychology, Complutense University of Madrid, Pozuelo de Alarcón, Madrid, Spain

[3] Electrical Engineering and Bioengineering Group, Department of Industrial Engineering & IUNE, University of La Laguna, San Cristobal de La Laguna, Tenerife, Spain

*Abstract:*

Despite the increase in calculation power in the last decades, the estimation of brain connectivity is still a tedious task. The high computational cost of the algorithms escalates with the square of the number of signals evaluated, usually in the range of thousands. In this work we propose a re-formulation of a widely used algorithm that allows the estimation of whole brain connectivity in much smaller times.

We start from the original implementation of Phase Locking Value (PLV) and re-formulated it in a highly computational efficient way. Besides, this formulation stresses its strong similarity with coherence, which we used to introduce two new metrics insensitive to zero lag synchronization, the imaginary part of PLV (iPLV) and its corrected counterpart (ciPLV).

The new implementation of PLV avoids some highly CPU-expensive operations, and achieved a 100-fold speedup over the original algorithm. The new derived metrics were highly robust in the presence of volume conduction. ciPLV, in particular, proved capable of ignoring zero-lag connectivity, while correctly estimating nonzero-lag connectivity.

Our implementation of PLV makes it possible to calculate whole-brain connectivity in much shorter times. The results of the simulations using ciPLV suggest that this metric is ideal to measure synchronization in the presence of volume conduction or source leakage effects.

## I. Introduction

The study of the brain caught the interest of researchers before neuroscience was even a discipline. For a long time, the brain was considered an organ, or rather a set of organs, with distinct parts taking care of distinct duties, following the so-called phrenological point of view. However, in recent years some have begun to departure from this traditional approach in which distinct parts of the brain are believed to have different functions. Instead, they argue than cognition, thought and action are supported by the collective action of sets of brain areas (networks) (Friston 1994). Within this paradigm, the brain areas form a network (the *connectome*), entwined by white matter tracts (anatomical connectome), in which different sub-networks communicate dynamically with each other to perform different functions (functional connectome).

Whereas the anatomical connectome is assessed from diffusion tensor imaging (DTI) (Le Bihan et al. 2001), the functional connectome is constructed from neuroimaging techniques such as functional magnetic resonance imaging (fMRI) and electro- and magneto-physiology (Friston et al. 2003; Brookes et al. 2011). In the latter case, however, the description of the connectome is not straightforward, as the underlying mechanics of the brain are unknown. Instead, we have to trust on measures of brain activity at different recording sites. In this framework, functional connectivity (FC) is defined as the existence of statistical dependence between the activities in two or more sites above chance level, a dependence that can be evaluated in different ways.

In the case of signals such as M/EEG, one of the most studied connectivity hypothesis is that of phase synchronization (PS) (Lachaux et al. 1999; Fries 2015; Garcés et al. 2016; López et al. 2014). Under this scenario, two brain areas show the activity of different oscillators and, in the case of connectivity between different regions, their oscillation properties (i.e. phase or frequency) should be related. If this relation can be mathematically evaluated, we get an estimator of phase connectivity.

Of the many PS measures available in the literature, one of the most used, mainly due to its simplicity, is the Phase Locking Value (PLV, (Lachaux et al. 1999)), sometimes called Mean Phase Coherence (MPC) (Mormann et al. 2000). This measure evaluates the instantaneous phase difference of the signals under the hypothesis that connected areas generate signals whose instantaneous phases evolve together. In this case, the phases of the signals are said to be "locked", and their difference is therefore constant. However, real-world signals are inherently noisy, and it is not always possible to be sure that the evaluated signal only comes from one oscillator.

The problem is solved by allowing some deviation from the condition of a constant phase difference. Thus, PLV evaluates the spread of the distribution of phase differences, and the connectivity estimation is linked to this spread. The narrower the distribution of the phase difference, the higher the PLV value, which ranges between zero (no phase dependence) and one (complete phase dependence). We will elaborate on the mathematics behind the estimation of the PLV in the next section. Here we briefly describe the main steps involved.

The most common method to calculate PLV is based on the instantaneous phase of the signals obtained using the Hilbert or the wavelet transforms. In both cases, the calculations are fast and reliable, which has undoubtedly contributed to its use. The calculation of the phase difference and its spread is also mathematically simple, and, in principle, not very time-consuming. However, even when the computation of PLV is fast and easy for a fair amount of data, the computational cost grows with the square of the number of signals. This is especially important in the case of M/EEG, distributed source-space analysis, where such number ranges from the thousands to the hundreds of thousands.

It is also noteworthy that, despite the estimation of the PLV and other related indices from data has a long history, from time to time new studies appear in the literature (Ewald et al. 2012; Kovach 2017) in which different aspects of the calculation procedures and the indices themselves are analyzed with a fresh eye to refine the existing methodologies. This paper intends to be one of such studies to shed new light on PS estimation, interpretation and use. Thus, we start from the original PLV formulation of Lachaux (Lachaux et al. 1999) and rewrite it to obtain an equivalent expression that is far easier to compute, reducing the required time for the PLV calculation up to a factor of 100. Besides, we also show that this new formulation is closely related to that of coherency, thereby allowing an interpretation of this well-studied function in terms of the PLV (see also (Kovach 2017)). As an additional advantage, such interpretation allows the formulation of two PLV-based zero-lag-insensitive measures: the imaginary PLV (iPLV) and the corrected iPLV (ciPLV), which can be used as alternatives to the imaginary part of the coherency (Nolte et al. 2004) and its corrected version (Ewald et al. 2012), respectively, in the assessment of direct PS from M/EEG data.

## II. Methods

### A. *Computational optimization*

In their original paper, Lachaux (Lachaux et al. 1999) defined the PLV as a time-dependent connectivity measured tailored to study evoked activity. The idea behind their definition is that the stimulus resets the phase of the neural oscillators so that signals connected in a given time should have a stable phase-difference along trials. Its mathematical formulation reads:

$$PLV_{i,j}(t) = \frac{1}{N}\left|\sum_{n=1}^{N} e^{-i\left(\varphi_i(t,n)-\varphi_j(t,n)\right)}\right| \tag{1}$$

where $N$ is the number of trials and $\varphi_i(t,n)$ is the instantaneous phase for signal $i$ in trial $n$ at time $t$.

This definition can be extended to resting-state data, by assessing phase locking as a stable phase-difference over time, thereby obtain the so-called MPC[1] (Mormann et al. 2000):

$$PLV_{i,j} = \frac{1}{T}\left|\sum_{t=1}^{T} e^{-i\left(\varphi_i(t)-\varphi_j(t)\right)}\right| \tag{2}$$

where $T$ is the data length.

In either case, one has to extract the instantaneous phase $\varphi(t)$ of each signal. Besides, for the phase to be physically meaningful, it is necessary that only one oscillator is present in each signal. This is achieved, e.g., by means of a narrow-band pass filtering or, equivalently, the convolution with a narrow band complex wavelet such as that of Morlet (Bruns 2004).

After the filtering process, we obtain a band-pass version of the Hilbert analytical signal:

$$cBP\{x(t)\} = x_{BP,H}(t) = x_{BP}(t) + i\tilde{x}_{BP}(t) = A(t)\cdot e^{-i\varphi_i(t)} \tag{3}$$

where $\tilde{x}$ represent the Hilbert transform of $x$, and BP stands for band pass. The instantaneous phase is the angle between the real and the imaginary parts of the Hilbert analytical signal, or the angle between the original (band-pass) signal and its Hilbert transform.

The instantaneous phase is usually extracted from this analytical signal, the phase difference estimated and, finally, the exponentiation calculated to get the unit phase difference vector. However, these two operations (phase extraction and exponentiation) are computationally expensive, but, as we will show, they can be easily circumvented by using the properties of exponentials.

First, let us obtain the oscillatory part of the analytical signal by normalizing (3):

$$\dot{x}_{BP,H,i}(t) = \frac{x_{BP,H,i}(t)}{\left|x_{BP,H,i}(t)\right|} = \frac{A_i(t)\cdot e^{-i\varphi_i(t)}}{A_i(t)} = e^{-i\varphi_i(t)} \tag{4}$$

From this, we easily derive the exponential of the phase difference:

---

[1] Henceforth, for the sake of simplicity, we will use $PLV_{i,j}$ to refer to both the time varying phase locking value as derived by Lachaux et al., and the mean phase coherence. In any case, both indices can be easily distinguished since the former one ($PLV_{i,j}(t)$) explicitly depends on time, whereas the latter one, averaged over the whole data segment, does not.

$$\dot{x}_{BP,H,i}(t) \cdot \left(\dot{x}_{BP,H,j}(t)\right)^* = e^{-i\varphi_i(t)} \cdot \left(e^{-i\varphi_j(t)}\right)^* = e^{-i\varphi_i(t)} \cdot e^{i\varphi_j(t)}$$

$$= e^{-i\left(\varphi_i(t)-\varphi_j(t)\right)} \tag{5}$$

where $(\cdot)^*$ represents complex conjugate. Thus, expression (2) can be rewritten as:

$$PLV_{i,j} = \frac{1}{T}\left|\sum_{t=1}^{T}\dot{x}_{BP,H,i}(t) \cdot \left(\dot{x}_{BP,H,j}(t)\right)^*\right| \tag{6}$$

or, using vector algebra:

$$PLV_{i,j} = \frac{1}{T}\left|\dot{x}_i \cdot \dot{x}_j^T\right| \tag{7}$$

where $\dot{x}_i$ is a vector version of $\dot{x}_{BP,H,i}(t)$ and $\dot{x}^T$, its transpose conjugate of $\dot{x}$. This calculation is computationally very efficient, and allows a considerable speed up of the estimation with low memory penalization. The calculation efficiency is discussed in the *Results* section.

The efficient formulation can be extended to Lachaux's $PLV_{i,j}(t)$ by constructing the vectors with the $t$-th sample of each trial:

$$PLV_{i,j}(t) = \frac{1}{T}\left|\dot{x}_{i,t} \cdot \dot{x}_{j,t}^T\right| \tag{8}$$

### B. Relation to coherence

If we expand (1) using the oscillatory part of $x$, we get:

$$PLV_{i,j}(t) = \frac{1}{N}\left|\sum_{n=1}^{N}e^{-i\varphi_i(t,n)} \cdot \left(e^{-i\varphi_j(t,n)}\right)^*\right| \tag{9}$$

Phase synchronization only makes sense for signals composed of a single oscillatory component. Yet, it is mathematically possible to calculate the PLV from both a broadband signal and an arbitrarily narrow band one, even when these calculations do not have, in principle, physical sense.

If we take the extreme case of a single oscillator whose spectrum is non-zero only at a given frequency, $f_o$, the signal can be written as an out-of-phase cosine, where the phase at the initial time is equal to the Fourier phase, and the phase at any other time is determined by the delay and the frequency.

$$x_f(t) = A \cdot \cos(2\pi f_o t + \varphi) = A \cdot \Re\{e^{-i(2\pi f_o t + \varphi)}\}$$
$$X_f(f_o) = \frac{A}{2} \cdot e^{-i\varphi} \tag{10}$$

where $X(f)$ is the Fourier transform of $x(t)$ at $f_o$ and $\Re\{x\}$ stands for the real part of $x$. In this case, Lachaux's definition of PLV is no longer time dependent, and its formulation can be simplified as:

$$PLV_{i,j} = \frac{1}{N}\left|\sum_{n=1}^{N} e^{-i\varphi_i(n)} \cdot \left(e^{-i\varphi_j(n)}\right)^*\right| = \frac{1}{N}\left|\sum_{n=1}^{N} \frac{X_i(f,n) \cdot \left(X_j(f,n)\right)^*}{|X_i(f,n)| \cdot |X_j(f,n)|}\right| \tag{11}$$

This formula closely resembles that of coherence (Nunez et al. 1997). In fact, coherence can be rewritten as:

$$COH_{i,j}(f) = \frac{\left|\sum_{n=1}^{N} X_i(f,n) \cdot \left(X_j(f,n)\right)^*\right|}{\sqrt{\sum_{n=1}^{N} |X_i(f,n)|^2 \cdot \sum_{n=1}^{N} |X_j(f,n)|^2}}$$
$$= \frac{\left|\sum_{n=1}^{N} |X_i(f,n)| \cdot |X_j(f,n)| \cdot \frac{X_i(f,n) \cdot \left(X_j(f,n)\right)^*}{|X_i(f,n)| \cdot |X_j(f,n)|}\right|}{\sqrt{\sum_{n=1}^{N} |X_i(f,n)|^2 \cdot \sum_{n=1}^{N} |X_j(f,n)|^2}} \tag{12}$$

With this formulation, it is clear that coherence is a weighted average of the unit phase vectors, i.e., a version of PLV weighted by the joint amplitude of the signals at a given frequency. Alternatively, coherence can be a regarded as a version of PLV weighted by the signal-to-noise (SNR) ratio of each trial, because if the environmental conditions (noise) are stable, the amplitude of a signal is proportional to its SNR. In any case, coherence and PLV are tightly related and both are different formulations of the same principle.

It can be argued that PLV and coherence are inherently different, as the PLV must be calculated over a whole oscillator (i.e. a frequency band) and coherence is calculated independently for each frequency. However, following the previous logic, it is trivial to understand that PLV in a given band is related to the average coherence in the frequency range encompassed by the band weighted by their relative amplitude. We will further elaborate on this relationship in the *Results* section.

### C. Zero-lag-insensitivity versions

Despite its popularity, the PLV presents an important limitation when used to assess functional brain connectivity: its sensitivity to volume conduction (Stam et al. 2007) and source-leakage effects. In fact, in the case of M/EEG data at the sensor level, several sensors can simultaneously pick up the activity from the same source (volume conduction). In turn, at the source level, due to the low spatial resolution of the data, different neighboring sources may share some activity (source leakage).

Luckily, due to the low capacitance of the tissues of the head for the physiological frequencies and the small distance that the currents have to travel, the propagation of the signals of interest can be considered instantaneous (Nolte et al. 2004; Stam et al. 2007). Under this assumption, volume conduction/source leakage occurs with zero-lag propagation. In other words, the phase difference of the part of the signals related to such spurious connectivity must be zero.

Following this logic, Cornelis Stam (Stam et al. 2007) and Guido Nolte (Nolte et al. 2004) came with two different PS metrics that discard zero-lag connectivity and are, thus, insensitive to volume conduction. They are the phase lag index (PLI) and the imaginary part of coherency, respectively. Due to the tight relation of PLV with coherency, it is possible to extend the imaginary part of coherency to PLV to obtain a PLV-based measure insensitive to volume conduction effects, which, for symmetry, we will term imaginary PLV (iPLV):

$$iPLV_{i,j,t} = \frac{1}{T}\Im\{\dot{x}_{i,t} \cdot \dot{x}_{j,t}^T\}$$

(13)

where $\Im\{x\}$ stands for the imaginary part of $x$. This measure is insensitive to zero-lag effects, as it removes the contribution of the zero phase differences that, due to the complex exponentiation, gives real PLV values. However, (13) is not normalized, as its upper bound, corresponding to two signals with a phase difference $\varphi>0$, is $\sin(\varphi)$. This can be corrected analogously to what (Ewald et al. 2012) did for the imaginary part of coherency, to define a corrected imaginary PLV (ciPLV):

$$ciPLV_{i,j,t} = \frac{\frac{1}{T}\Im\{\dot{x}_{i,t} \cdot \dot{x}_{j,t}^T\}}{\sqrt{1 - \left(\frac{1}{T}\Re\{\dot{x}_{i,t} \cdot \dot{x}_{j,t}^T\}\right)^2}}$$

(14)

This definition of ciPLV is similar to the definition of lagged coherence as introduced by Pascual-Marqui and coworkers (Pascual-Marqui et al. 2011).

## III. Results

### A. Speedup achieved using the proposed algorithm

In order to study the behavior of the proposed formulation of the PLV algorithm, we calculated PLV using three different implementations for different data lengths and number of signals. Even when the computation is completely deterministic, and any data set with the same characteristics would produce the same results, for the sake of fidelity, we used real source space data.

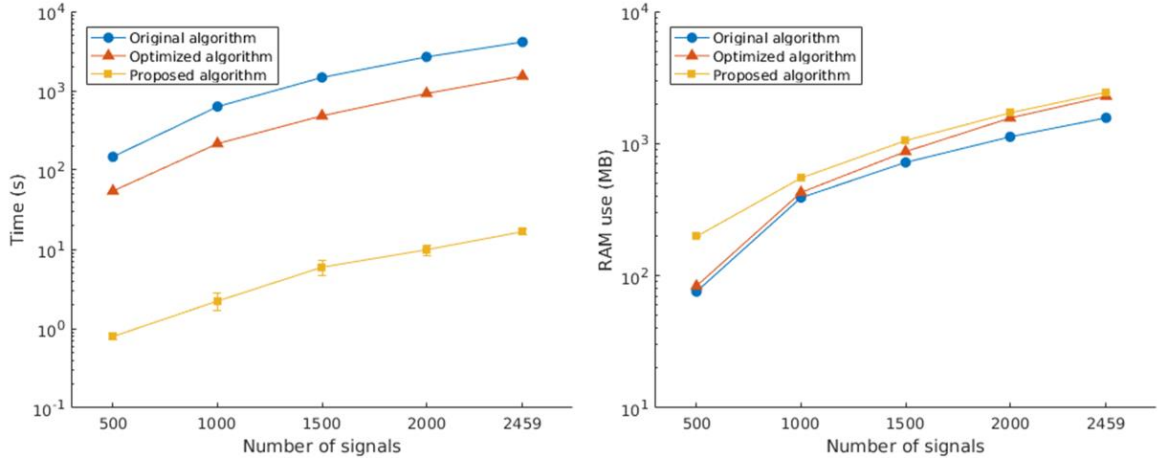

Figure 1. Execution time (left) and RAM use (right) of the three evaluated PLV algorithms for a different number of signals and 40 trials. The plotted values for each data point are mean and standard deviation of the execution time and RAM use over 5 executions. The proposed algorithm achieves a 100-fold speed-up with a small increase in memory use. Please note the logarithmic scale of the y-axis.

Test data consisted of five minutes of eyes-closed resting-state magnetoencephalographic (MEG) activity acquired using a 306-sensors Elekta Vectorview system (Elekta, AB, Stockholm, Sweden), located inside a magnetically shielded room (VacuumSchmelze GmbH, Hanau, Germany) at the Laboratory for Cognitive and Computational Neuroscience (Madrid, Spain). Data was acquired with a sampling rate of 1000 Hz (anti-alias band-pass filter of 0.1-330 Hz) and filtered using a spatiotemporal signal space separation method (Taulu & Simola 2006).

Data were segmented in 20 or 40 four-second artifact-free segments and band-pass filtered in the classical alpha band (8 to 12 Hz) using a 2000th order FIR filter in two passes with 2 seconds of real data as padding on both sides. The instantaneous phase of the signal was determined using Hilbert's analytical signal. In order to avoid edge effects, the analytical signal was calculated prior to the removal of the padding. In order to get usable timescales for the original implementations, data was downsampled by a factor of ten.

We then placed 2459 dipoles inside the head of the subject, in a 1 cm homogeneous three-dimensional grid. Source space data was calculated using a realistic single shell as a forward model (Nolte 2003) and a beamformer as inverse method (van Veen et al. 1997). The obtained spatial filter was applied to the sensor space data to obtain up to 2459 time series in source space. Last, all-to-all PLV connectivity matrix was calculated for different numbers of sources, ranging from 500 to 2459. PLV calculation was performed using Matlab (The Mathworks Inc., Natick, Massachusetts) and the three functions whose codes are in the *Appendix*. Time and memory performances were measured using Matlab's tic/toc and FieldTrip's (Oostenveld et al. 2011) memtic/memtoc functions, respectively.

Figure 1 shows the results of the performance test using 40 trials and different numbers of signals. Every configuration was evaluated 5 times, and the figure shows the mean and the standard deviation. The original implementation is the slowest, but the most memory-efficient of the three evaluated algorithms. The optimized implementation achieves almost a 3-fold speedup with a slight increase in memory consumption. However, the proposed algorithm is able to get a 100-fold speedup over the optimized implementation, with only a marginal (especially at a high number of signals) increase in memory use. Results using 20 trials of data are not shown here, but the behavior was similar, with approximately half the time and memory consumption.

These results show that the proposed algorithm can achieve a dramatic speed-up in the calculation of the PLV values. The algorithm even allows the calculation of PLV from non-downsampled data, calculating the full connectivity matrix of 2459 time-series in an average of 113 seconds, using an average of around 8 GB of RAM memory, over 100 executions. Extrapolating the results shown in the previous paragraph, this execution, even with the optimized implementation, would take 3 hours.

### B. Comparison of PLV and coherence

As noted in the *Methods* section, both PLV and coherence are phase-synchronization estimation algorithms. Both metrics measure similar properties of the data, but while coherence gives one result per frequency, PLV gives one result per frequency band. This difference arises from the fact that coherence estimates phase synchronization from Fourier's phase, whereas PLV estimates it from Hilbert's phase. In the extreme case of an infinitely narrow band, both phase definitions converge, but in a normal case where the band is finite, the results of coherence and PLV would diverge.

In order to evaluate the behavior of the coherence and the PLV in signals with a known degree of coupling, we used a pair of coupled chaotic systems: a Rössler system and a Lorenz one (Quian-Quiroga et al. 2000). In this setup, the Rössler system acts as driver, and an oscillation frequency can be defined from it. The slave Lorenz system is driven by the Rössler to an extent determined by the coupling parameter $C$, ranging from zero (completely independent systems) to one. The mathematical definition of the coupled systems is:

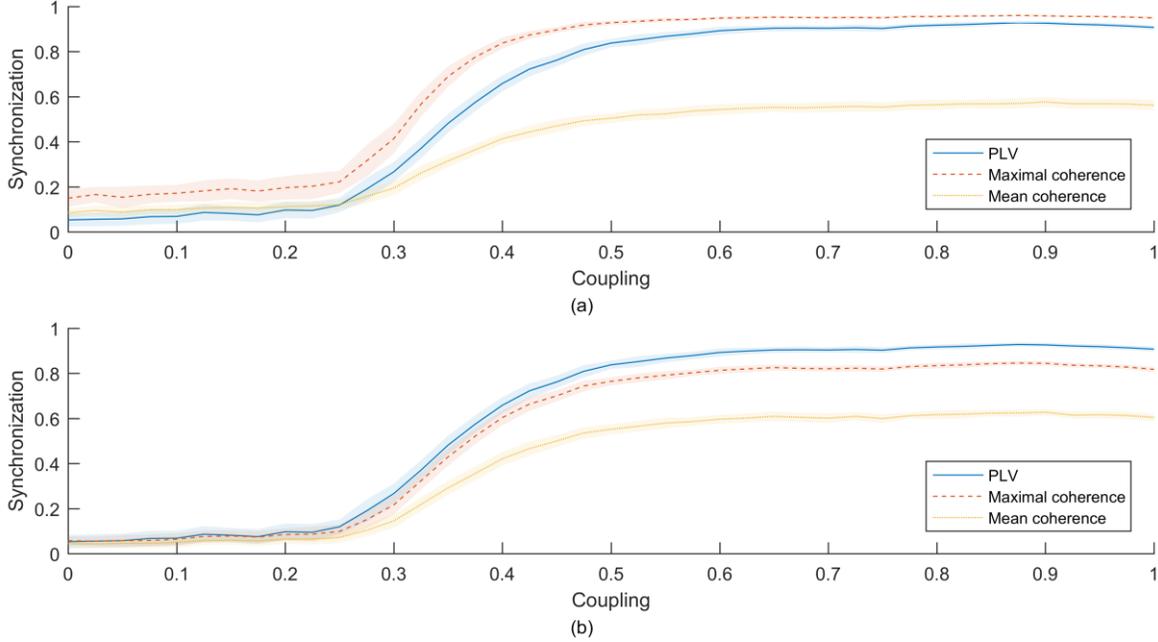

Figure 2. Values of the synchronization indices analyzed for different couplings for a pair of Rössler and Lorenz systems. The dark line represents the mean value over 50 executions, and the shadow line represents the standard deviation. (a) PLV, maximal and average coherence for the band between $0.570\bullet\pi$ and $0.600\bullet\pi$ radians per sample. Coherence was calculated using a Hamming window of length 400 with 200 samples of overlapping. (b) The same that in (a), using a window length of 100 samples with 50 samples of overlapping. Note that only coherence-based metrics are affected by this change.

$$
\begin{cases}
\dot{x_1} = -a \cdot (y_1 + z_1) \\
\dot{y_1} = a \cdot (x_1 + 0.2 \cdot y_1) \\
\dot{z_1} = a \cdot (0.2 + z_1 \cdot x_1 - 5.7 \cdot x_1)
\end{cases}
$$
$$
\begin{cases}
\dot{x_2} = 10 \cdot (y_2 - x_2) \\
\dot{y_2} = 28 \cdot (x_2 - y_2 - x_2 * z_2 + C \cdot y_1{}^2) \\
\dot{z_2} = x_2 \cdot y_2 - {}^8\!/_3 \cdot z_2
\end{cases}
\tag{15}
$$

where subscripts 1 and 2 refer to the Rössler and the Lorentz system, respectively, $a$ is a scaling parameter determining the fundamental frequency of the Rössler oscillator and $C$ is the coupling parameter. The scaling parameter was set to 10, establishing the oscillatory frequency at $0.585\cdot\pi$ radians per sample, and the oscillatory band was set to $0.570\cdot\pi$ to $0.600\cdot\pi$ radians per sample. With this setup, we generated 50 pairs of signals of 20,000 samples for different values of $C$ ranging between zero and one. PS between the systems was finally estimated using the first variable of each chaotic system $x$.

In order to get a band-wise coherence value, we adopted two different approaches using either the maximum or the mean value of coherence over the evaluated band. The values of PLV and both coherence approaches are shown in Figure 2a. Clearly, both PLV and coherence evolve closely with increasing coupling, with coherence values slightly overestimated for low couplings. This overestimation is due to the lower number of phases used for the coherence calculation as compared to the PLV. Indeed, for the calculation of coherence data was split into 400 samples segments with 200 samples of overlapping, giving a total of 99 segments, and 99 independent phases. For the calculation of PLV, with a bandwidth of $0.030 \cdot \pi$ radians per sample, there is three effective sample for every 100 samples, giving a total of 600 effective samples, and thus 600 independent phases.

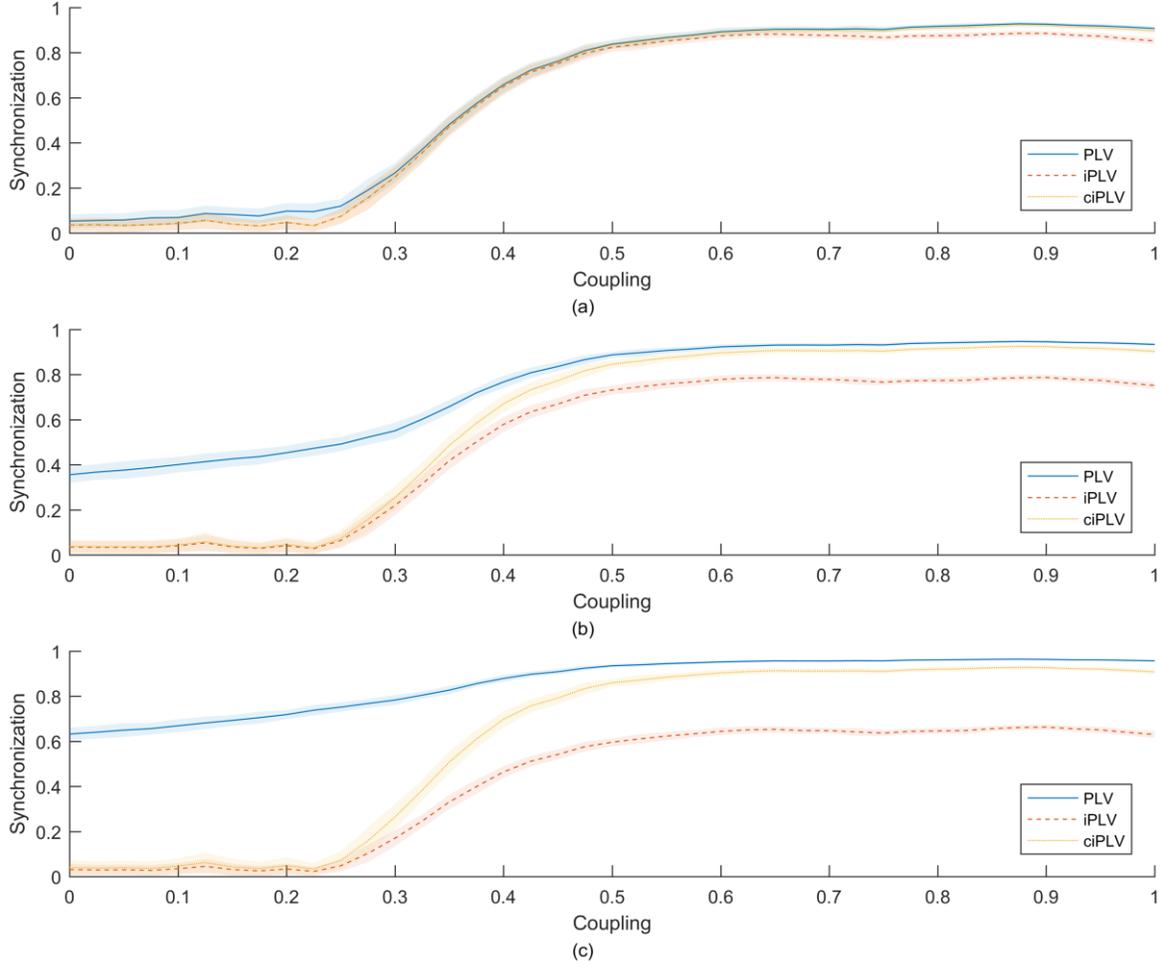

Figure 3. Values of the synchronization indices analyzed for different couplings for one Rössler and one Lorenz system. The dark line represents the mean value over 50 executions, and the shadow line represents the standard deviation. (a) Synchronization estimated using PLV, iPLV and ciPLV. (b) The same that in (a), after adding a 10% of linear mixing to the signals. (c) Idem after adding a 20% of linear mixing.

To address this problem, Figure 2b shows the results using a smaller window for the coherence calculation. In this case, with 100 samples windows and 50 samples overlap, coherence is calculated using 399 phases. The overestimation for low coupling is clearly lower than in the previous case, but the maximal value of coherence drops to 0.85. In this case, the reduction in the calculated synchronization can be explained from the frequency smoothing associated with a shorter time window. Due to the narrowness of the frequency peak of the Rössler oscillator, a larger amount of frequency smoothing forces the coherence to be determined with a phase that includes both the peak itself and the surrounding frequencies.

The effect is similar when evaluating the mean coherence over a band, as coupled frequencies are averaged together with non-coupled ones. However, PLV does not show this effect, as the instantaneous phase of the signal is calculated over the whole band, and not from an average of frequencies. In this sense, PLV is superior to coherence, as the position and narrowness of the oscillator over the observed band are not relevant. PLV could even handle an oscillator with varying frequency, given that the oscillatory frequency never goes outside of the observed band. On the other hand, maximal coherence would only return the coupling at the most common frequency, ignoring the coupling at other frequencies, and mean coherence would dramatically underestimate the connectivity.

### C. Effects of volume conduction

As commented above, one of the main criticisms to PLV and coherence as applied to brain signals is their sensitivity to volume conduction. It can be modeled as an instantaneous projection of one signal onto the other, giving rise to zero-lag synchronization. As shown in the Methods section, Guido Nolte's proposed imaginary part of coherency (Ewald et al. 2012; Nolte et al. 2004) can be extended to the PLV, giving a set of zero-lag insensitive measures. In order to evaluate the behavior of these PLV derived metrics, we used the same pair of chaotic systems defined in the previous section. The systems show nonzero-lag phase synchronization, and the volume conduction can be introduced using instantaneous linear mixing (Porz et al. 2014; Haufe et al. 2013):

$$\tilde{x} = x + V \cdot y$$
$$\tilde{y} = y + V \cdot x \tag{16}$$

where $V$ is a parameter determining the amount of mixing. Figure 3 shows the synchronization estimated between the signals described in the previous section using the PLV, iPLV and ciPLV, and values of $V$ of 0, 0.1 and 0.2.

Figure 3a shows the estimated synchronization for V=0 (without zero lag mixing), measured with PLV, iPLV and ciPLV. Values of PLV and ciPLV are almost identical, indicating that the synchronization of the pair of the chaotic oscillator is nonzero lag. However, iPLV values are slightly lower, especially for high couplings, indicating that the lag between both signals is almost, but not exactly, one-quarter of a cycle, so the PLV complex vector has only a small real component.

Figure 3b, corresponding to V=0.1, shows the effect of 10% volume conduction between the signals, introducing low spurious zero-lag synchronization. PLV values increase especially at lower couplings, whereas iPLV and ciPLV values remain unaffected. iPLV shows lower synchronization values that ciPLV, which remains almost identical to the V=0 case. This is because the complex PLV vector now has an important real component that is completely ignored by iPLV. This shows that ciPLV is superior, as it uses this real component to normalize the imaginary one. Figure 3c shows the effect of 20% volume conduction, with similar results, but exacerbating the estimation errors in PLV and iPLV. ciPLV, however, continues to extract the correct synchronization value.

## IV. Discussion

The formulation of PLV described in this paper allows for a speedup of two orders of magnitude in the calculation of this index. This result is especially relevant in the study of EEG/MEG source-space connectivity. Source space models based in volumes, as the one used in the *Results* section above, include around 2500 sources for the whole brain, and around 1500 for gray matter. In the case of sources based in surfaces, which are normally used in minimum norm estimates (Pascual-Marqui et al. 1994; Gramfort et al. 2013), the number of sources ranges between 1000 and 10000 per hemisphere. In both cases, a whole brain connectivity study, even with subsampling, would take up to several hours per subject (and band).

The usual approach in those cases is the parcellation of the brain and the extraction of a representative time course per parcel (Farahibozorg et al. 2017; Korhonen et al. 2014). This reduces the number of time courses to the order of the hundreds, thereby allowing the estimation of whole-brain connectivity in a reasonable amount of time. However, this method forces the definition of homogeneous parcels, defined by one unique time series, thus discarding some information associated to inter-area variability. With the proposed algorithm the calculation of source-to-source PLV would take only a few minutes per subject (and band), allowing the estimation of whole-brain connectivity for a whole study in a matter of hours.

In addition, our reformulation of the original PLV definition stressed the similarities between PLV and coherence. The comparison of both metrics as applied to a pair of chaotic systems with different coupling levels showed that PLV and coherence values are closely related, with PLV showing a better behavior in the (very realistic) situation where the coupling takes place over a whole (possibly narrow) band instead of at a single frequency. This is, however, not a flaw of coherence, but the direct effect of its definition. PLV seems best to evaluate synchronization over a whole band, whereas coherence does to evaluate it at fixed frequencies.

Finally, we evaluated the behavior of PLV in the relevant case of volume conduction/source leakage as simulated by instantaneous linear mixing of the signals corresponding to the two systems. In analogy to coherency, we introduced the imaginary part of PLV and its corrected version. Both algorithms proved insensitive to volume conduction, but iPLV was flawed by the effect of a real component of the complex PLV vector. As for the imaginary part of coherency, iPLV only reaches the maximal value when the phase difference of the two signals is exactly $\pi/2$ radians, whereas it drops to almost zero when the phase difference is small but consistently nonzero. ciPLV corrects this behavior in a similar way to the corrected imaginary part of coherency (Ewald et al. 2012), being both unbiased and insensitive to volume conduction/source leakage effects. This is not the first time that a metric based on the imaginary part of PLV is used (Dimitriadis et al. 2017; Palva et al. 2017; Wang et al. 2018), but, to our knowledge, the expected behavior had not been thoroughly tested yet.

Both sensor and source space EEG/MEG connectivity have to deal with the burden of zero-lag connectivity. Zero lag connectivity between two sensors is usually interpreted as the effect of a common source in MEG and an effect of volume conduction in EEG. At the source level, it is usually interpreted as an effect of source leakage. Both interpretations are based in the idea that brain signals cannot travel instantaneously between different parts of the brain, and any real connectivity should imply a delay (Stam et al. 2007). Even when some scenarios allow zero-lag connectivity in the brain (Kovach 2017), it is not possible to distinguish real zero-lag connectivity from volume conduction or source leakage.

Several measures have been developed to try to overcome this problem. The most notable of these measures is the imaginary part of coherence, and its corrected counterpart. However, even when coherence is effectively a phase-synchronization measure, the results do not completely fit in the phase synchronization hypothesis in the brain. PLI is another measure developed to estimate phase connectivity ignoring the contribution of zero lag (Stam et al. 2007), but the metric has showed a low test-retest reliability (Garcés et al. 2016; Colclough et al. 2016).

The variations of PLV proposed in this paper mimic their coherence counterparts, and have proved effective to remove zero-lag connectivity while keeping intact nonzero-lag connectivity. This allows the study of sensor or source space EEG/MEG connectivity evaluating only real synchronizations, and completely ignoring volume conduction or source leakage ghost synchronizations. This method, as discussed, is not perfect (Kovach 2017), but in the same fashion that imaginary coherence (Nolte et al. 2004), allows the evaluation of real connectivity without the influence of zero-lag interference.

Note, however, that in any case, and according to the latest results in the literature (Palva et al., 2017; Wang et al., 2017) no bivariate FC index (whether zero lag insensitive or not) is free from the effect of spurious detection of connectivity due to source leakage. Nevertheless, for the detection of true connectivity, we believe that both iPLV and especially ciPLV are excellent choices that can be very efficiently estimated using the algorithms we introduced here.

## V. Conclusion

In this paper, we propose a new formulation of the PLV algorithm that allows its fast computation by bypassing CPU-demanding calculations. The direct use of this new algorithm in Matlab allows for a speedup by a factor of 100 from a vector-optimized implementation. Even though this implementation can be improved, it is unlikely that a similar speedup can be achieved using the original formulation.

Noteworthy, the optimization is only based in a mathematical reformulation of the original definition of PLV, in which the phase extraction and the exponentiation necessary for its calculation are replaced by a simple matrix multiplication. Besides, the algorithm can be implemented in any language, and the code provided in the *Appendix*, created in Matlab, is completely machine-independent. Thus, differently to a toolbox we recently released (García-Prieto et al. 2017), there is no need to compile the source code in C language, which makes the present formulation much easier to use. Note also that this paper only deals with the mathematical definition, and there may be room for computational improvement.

In conclusion, we have shown that, despite its longevity, it is still possible to refine further the estimation of the (allegedly) most popular method of PS, the PLV index. The new formulation not only allows for a much faster estimation of the index but also, by highlighting the similarities between PLV and coherence, prompts the definition of two new measures insensitive to zero lag. These measures, derived from the vector product of the real and imaginary part of PLV, are analog to the (corrected) imaginary part of coherence and are therefore robust again volume conduction and source leakage effects. We hope these new developments will encourage further the use PLV-related measures in the analysis of neurophysiological data at the sensor and the source level.

## Appendix

In the Results section, we tested three different implementations for the calculation of PLV. In this Appendix, we show the three different codes developed in Matlab to evaluate the behavior of each algorithm. In the codes provided, all three implementations are fed with the Hilbert analytic signal, with shape signals per samples per trials.

The first implementation includes two nested loops, so every pair of signals is evaluated independently. This option is the most memory conservative but also the slowest one, as only two signals are evaluated in each interaction. The code is:

```
[ nc, ns, nt ] = size( data );
phs  = angle( data );

plv  = zeros( nc, nc, nt );
for t = 1: nt
    for c1 = 1: nc
        for c2 = c1 + 1: nc
            dphs = phs( c1, :, t ) - phs( c2, :, t );
            plv( c1, c2, t ) = abs( mean( exp( 1i * dphs ) ) );
        end
    end
end
```

The second implementation is a vectorized and memory efficient version of the algorithm. In this implementation every signal is compared against all the other signals, avoiding two loops. This option uses slightly more memory but achieves a speedup of factor 2.5 over the original one. The code is:

```
[ nc, ns, nt ] = size( data );
phs  = angle( data );

plv  = zeros( nc, nc, nt );
for t = 1: nt
    tplv = complex( zeros( nc ) );
    for s = 1: ns
        dphs = bsxfun( @minus, phs( :, s, t ), phs( :, s, t )' );
        tplv = tplv + exp( 1i * dphs );
    end
    plv( :, :, t ) = abs( tplv / ns );
end
```

The last implementation uses the proposed algorithm. This algorithm allows the direct calculation of all source pairs at once, with negligible memory increase. In addition to avoiding the angle and exp functions, the implementation removes the need to use a loop over the signals. The code is:

```
[ nc, ns, nt ] = size( data );
ndat = data ./ abs( data );

plv  = zeros( nc, nc, nt );
for t = 1: nt
```

```
    plv( :, :, t ) = abs( ndat( :, :, t ) * ndat( :, :, t )' ) / ns;
end
```

**Acknowledgment**

The authors would like to thank X. Bello for her review of the style of this manuscript. This study was partially supported by one project from the Spanish Ministry of Economy and Competitiveness (TEC2016-80063-C3-2-R) to EP and one pre-doctoral fellowship from the Spanish Ministry of Education (FPU13/06009) to RB.